\documentclass[12pt,oneside]{article}
\usepackage{amsmath,amssymb,amsfonts,amsthm}
\usepackage{color}
\usepackage{multirow}
\usepackage{rotating}
\usepackage{paralist}
\usepackage{centernot}
\usepackage{stmaryrd}
\usepackage[retainorgcmds]{IEEEtrantools}
\newcommand{\dd }{\displaystyle}

\newcommand{\bq }{\begin{equation}}
\newcommand{\eq }{\end{equation}}

\newcommand{\bt}{\begin{theorem}}
\newcommand{\et}{\end{theorem}}

\newcommand\undersym[2]{\raisebox{-6pt}{\tiny$#2$}{\kern-5pt}\mbox{$#1$}}
\newcommand\overcirc[1]{\raisebox{10pt}{\tiny$\circ$}{\kern-6.5pt}\mbox{$#1$}}

\newtheorem{thm}{Theorem}[section]

\newtheorem{lem}[thm]{Lemma}

\newtheorem{defn}[thm]{Definition}

\newtheorem{rem}[thm]{Remark}

\numberwithin{equation}{section}

\begin{document}
\title{{\bf Gravity theory in SAP-geometry}\footnote{arXiv number: 1407.0938 [gr-qc]} }
\author{{\bf{Nabil L. Youssef}, Amr M. Sid-Ahmed and {Ebtsam H. Taha}}}
\date{}

\maketitle                     
\vspace{-1cm}
\begin{center}
Department of Mathematics, Faculty of Science,\\ Cairo
University, Giza, Egypt\\

\vspace{10pt}

nlyoussef@sci.cu.edu.eg,\, nlyoussef2003@yahoo.fr\\
amr@sci.cu.edu.eg,\, amrsidahmed@gmail.com\\
ebtsam.taha@sci.cu.edu.eg,\, ebtsam.h.taha@hotmail.com
\end{center}

\vspace{0.5cm} \maketitle
\smallskip

\noindent{\bf Abstract.}  The aim of the present paper is to construct a field theory in the context of absolute parallelism (Teleparallel) geometry under the assumption that the canonical (Weitzenb\"{o}ch) connection is semi-symmetric. The field equations are formulated using a suitable Lagrangian first proposed by Mikhail and Wanas. The mathematical and physical consequences arising from the obtained field equations are investigated.





\section{Introduction and Motivation }

Einstein spent the last years of his life trying
to build a geometric field theory in two main series of trails, to unify gravity and electromagnetism, known in the literature as
\lq\lq Unified Field Theories\rq\rq.\, In the first of these attempts, \lq\lq Einstein's absolute parallelism theory\rq\rq, he used the Absolute Parallelism geometry (AP-geometry). In his second attempt, \lq\lq Einstein's non-symmetric theory\rq\rq,\, he used another type of non-symmetric geometry \cite{AE}. Unfortunately, all these attempts were unsuccessful or incomplete.
This quest of unification preoccupied Einstein in vain during the last decades of his life as he tried to modify his basic equations of general relativity in an attempt to make additional room within the geometry of space-time for matter and force.

\smallskip
One of the successful attempts providing a unification of gravity and electromagnetism was accomplished by Mikhail and Wanas \cite{Mikhail}. The theory was formulated in the context of AP-geometry. Unlike Riemannian geometry, which has only ten degrees of freedom (in dimensional 4) just enough to describe gravity, AP-geometry has sixteen degrees of freedom. These extra degrees of freedom make AP-geometry a suitable mathematical framework for describing gravity and electromagnetism on geometric basis. It should be noted that this approach may be thought of as another alternative to the idea of
increasing the dimension of the underlying manifold as in the Kaluza-Klein theory.

\smallskip
In this paper we construct a set of field equations in the context of AP-geometry under the
additional assumption that the canonical (Weitzenb\"{o}ch) connection is semi-symmetric. We refer to this space as SAP space. We will use the same Lagrangian applied in the Generalized Field Theory (GFT) of Mikhail-Wanas. The reason for this choice will be clarified later on. As expected, SAP-space, being subject to more restrictions, reveals physical properties that do not necessarily hold in the general context of AP-space.
Moreover, many relations obtained acquire a much simpler and tangible form than their counterparts in AP-geometry. In particular, our chosen Lagrangian, which is similar in form to its counterpart in the GFT, acquires a much
simpler form. This, in turn, largely simplifies the calculations and gives rise to some interesting and unexpected results.

\smallskip
The paper is organized as follws. Section
$1$ provides a brief account of the basic concepts of AP-geometry.  Section $ 2$ gives a short survey on the notion of a semi-symmetric connection. Section $3$ gives a brief and concise outline of the GFT. In section $4$, variational principle is applied on our chosen Lagrangian, under the assumption that the canonical connection is semi-symmetric, and physical consequences of the obtained field equations are discussed. Section 5 deals with a comparative study between AP-geometry and GFT field equations, on one side, and SAP-geometry and our field equations, on the other side. The paper is ended with some comments and concluding remarks (Secion 6).


\section{Absolute Parallelism geometry}

Absolute parallelism geometry has gained more attention in recent years in constructing and applying modified gravity theories (such as
TEGR, cf., \cite{Nashed3}) 
and $f(T)$-theories (cf., \cite{Nashed1}, \cite{Nashed2}, \cite{Nashed4}).

In this section, we give a short survey of the absolute parallelism geometry or the geometry of parallelizable manifolds. For more details, we refer, for example, to \cite{W}, \cite{YE} and \cite{AMR}.\vspace{-5pt}
\begin{defn} A parallelizable manifold is an $n$-dimensional smooth
manifold $M$ which admits $n$ independent vector fields
$\,\undersym{\lambda}{i}$ $(i = 1, ..., n)$ defined globally on $M$.
\end{defn}
\vspace{-5pt}
This space is also known in the literature as Absolute Parallelism space (AP-space) or teleparallel space.
Let $ \,\undersym{\lambda}{i}^{\mu} \,\,\, (\mu = 1,2,...,n) $ be
the coordinate components of the \emph{i-th} vector field $\,\undersym{\lambda}{i}$. The Einstein summation convention is
applied to both Latin (mesh) and Greek (world) indices, where all
Latin indices are written beneath the symbols. The covariant components\, $\undersym{\lambda}{i}_{\mu}$\, of\, $\undersym{\lambda}{i}$\, are given by the relations
\begin{equation*}\label{covariant components of lambda}
\undersym{\lambda}{i}^{\mu} \,\, \undersym{\lambda}{i}_{\nu} = \delta^{\mu}_{\nu},    \quad       \undersym{\lambda}{i}^{\mu} \,\, \undersym{\lambda}{j}_{ \mu} = \delta_{i j}.
\end{equation*}

The canonical (or Weitzenb\"{o}ch) connection $\Gamma^{\alpha}_{\mu\nu}$ is defined by
\begin{equation}\label{canonical}
\Gamma^{\alpha}_{\mu \nu} := \undersym{\lambda}{i}^{\alpha} \,\, \undersym{\lambda}{i}_{ \mu ,\nu},
\end{equation}
where the comma here denotes partial
differentiation with respect to the coordinate function $x^{\nu}$. As easily checked, we have
\begin{equation*}\label{The condition of absolute parallelism}
 \lambda_{\mu | \nu} = 0,       \quad         {\lambda^{\mu}}_{|\nu} = 0,
\end{equation*}
where the stroke \lq \lq $|$" denotes covariant differentiation with respect to the canonical connection\, $\Gamma^{\alpha}_{\mu \nu}$.
The torsion tensor $\Lambda^{\alpha}_{\mu\nu}$ of $\Gamma^{\alpha}_{\mu \nu}$ is given as usual by
\begin{equation}\label{tor}
\Lambda^{\alpha}_{\mu \nu} := \Gamma^{\alpha}_{\mu \nu} - \Gamma^{\alpha}_{\nu \mu}\end{equation}

On the other hand, the curvature tensor $R^{\alpha}_{\epsilon \mu \nu}$ of the canonical connection  $\Gamma^{\alpha}_{\mu \nu}$ vanishes identically.
Hence, the AP-space is flat with respect to the canonical connection. However, there are
other three \emph{natural connections} which are non-flat. Namely, the dual connection
$\widetilde{\Gamma }^{\alpha}_{\mu \nu} := \Gamma^{\alpha}_{\nu \mu},$ the symmetric connection
$\widehat{\Gamma}^{\alpha}_{\mu \nu} := \frac{1}{2} (\Gamma^{\alpha}_{\mu \nu} + \Gamma^{\alpha}_{\nu \mu}) =\Gamma^{\alpha}_{(\mu \nu)}$
 and the Levi-Civita connection
 \begin{equation}\label{Riemannian AP}
 \overcirc{\Gamma}{^{\alpha}_{\mu \nu}} := \frac{1}{2} g^{\alpha \epsilon} (g_{\epsilon \nu ,\mu} + g_{\epsilon \mu , \nu} - g_{\mu \nu , \epsilon})
  \end{equation}
  associated with the metric structure defined by
  \begin{equation}\label{metric AP}
    g_{\mu \nu} := \undersym{\lambda}{i}_{\mu} \,\, \undersym{\lambda}{i}_{\nu}.
  \end{equation}
 The covariant derivatives with respect to the dual, symmetric and Levi-Civita connection will be denoted by $\widetilde{|},\,\widehat{|},\, \text{and} \, ;$ respectively.

The contortion tensor is defined by anyone of the following equivalent formulae

\begin{equation}\label{cont}
\gamma^{\alpha}_{\mu\nu}:=\, \undersym{\lambda}{i}^{\alpha}\, \undersym{\lambda}{i}_{\mu;\,\nu}, \quad\gamma^{\alpha}_{\mu\nu}= \Gamma^{\alpha}_{\mu\nu}-\, \overcirc{\Gamma}{^{\alpha}_{\mu\nu}}.
\end{equation}
Since $\, \overcirc{\Gamma}{^{\alpha}_{\mu\nu}}$ is symmetric, it follows that
\begin{equation*}
\label{v}\Lambda^{\alpha}_{\mu\nu}= \gamma^{\alpha}_{\mu\nu}- \gamma^{\alpha}_{\nu\mu}.
\end{equation*}
Furthermore, the basic form $C_{\mu}$ is defined by
\begin{equation}
\label{basic} C_{\mu} : = \Lambda^{\alpha}_{\mu\alpha} = \gamma^{\alpha}_{\mu\alpha}.
\end{equation}

\vspace{8pt}
Table 1 summarizes the geometry of the AP-space \cite{AMR}.

\begin{center} Table 1: Geometry of the AP-space\\[0.2cm]
{\begin{tabular}
{|c|c|c|c|c|c|c|c|c|c|c|c|}\hline
 \multirow{2}{*} {Connection}&\multirow{2}{*} {Coefficients}&Covariant
 &\multirow{2}{*} {Torsion}&\multirow{2}{*} {Curvature}&\multirow{2}{*} {Metricity}\\
 && derivative&&&\\[0.2cm]\hline
  Canonical&$\Gamma^{\alpha}_{\mu\nu}$&$|$
 &$\Lambda^{\alpha}_{\mu\, \nu}$&0&metric\\[0.2cm]\hline
 Dual&$\widetilde{\Gamma}^{\alpha}_{\mu\nu}$&$\begin{array}{cc}\tilde {}\\[-0.3cm]|\end{array}$
 &$-\Lambda^{\alpha}_{\mu\, \nu}$&$\widetilde{R}^{\alpha}_{\mu\nu\sigma}$&non-metric
 \\[0.2cm]\hline
 Symmetric&$\widehat{\Gamma}^{\alpha}_{\mu\nu}$&$\begin{array}{cc}
\hat {}\\[-0.3cm]|\end{array}$&0&$\widehat{R}^{\alpha}_{\mu\nu\sigma}$&non-metric\\[0.2cm]\hline
 Levi-Civita&$\overcirc{\Gamma}{^{\alpha}_{\mu\nu}}$&$;$&0&$\overcirc{R}{^{\alpha}_{\mu\nu\sigma}}$&metric\\[0.2cm]\hline
\end{tabular}}
\end{center}
\vspace{0.4 cm}

Table 2 gives a list of the most important second rank tensor fields of AP-geometry which play a key role in physical applications
(cf. \cite{Mikhail}, \cite{W1}, \cite{W2}). Moreover, most second rank tensor fields which have physical
significance in the AP-context can be expressed in terms
of these tensor fields. This table was first constructed by Mikhail \cite{FI}.
\vspace{0.05cm}
{\begin{center}{ Table 2: Fundamental second rank tensor fields of the AP-space}\\[0.2 cm]\
{\begin{tabular}{|c|c|}\hline
 Skew-Symmetric& Symmetric
\\[0.2 cm]\hline
$\xi_{\mu\nu}: = \gamma_{\mu\nu}  \!^{\alpha} \!\,_{|\alpha}$&
\\[0.2 cm]\hline
$\gamma_{\mu\nu}: = C_{\alpha}\gamma_{\mu\nu} \!^{\alpha}$&
\\[0.2 cm]\hline
$\eta_{\mu\nu} := C_{\alpha}\,\Lambda^{\alpha}_{\mu\nu}$&
$\phi_{\mu\nu} := C_{\alpha}\,(\gamma^{\alpha}_{\mu\nu}+\gamma^{\alpha}_{\nu\mu})$
\\[0.2 cm]\hline
$\chi_{\mu\nu} := \Lambda^{\alpha}_{\mu\nu|\alpha}$& $\psi_{\mu\nu}
:= \gamma^{\alpha}_{\mu\nu|\alpha}+\gamma^{\alpha}_{\nu\mu|\alpha}$\\[0.2 cm]\hline
$\epsilon_{\mu\nu} := C_{\mu|\nu} -
C_{\nu|\mu}$& $\theta_{\mu\nu} :=  C_{\mu|\nu}
+ C_{\nu|\mu}$\\[0.1 cm]\hline
{$\kappa_{\mu\nu} :=
\gamma^{\sigma}_{\alpha\mu}\gamma^{\alpha}_{\nu\sigma}
 - \gamma^{\sigma}_{\mu\alpha}\gamma^{\alpha}_{\sigma\nu}$}&
{$\varpi_{\mu\nu}: =
\gamma^{\sigma}_{\alpha\mu}\gamma^{\alpha}_{\nu\sigma} +
\gamma^{\sigma}_{\mu\alpha}\gamma^{\alpha}_{\sigma\nu}$}
\\[0.2 cm]\hline
$ \ $&$\sigma_{\mu\nu} :=
\gamma^{\sigma}_{\alpha\mu}\gamma^{\alpha}_{\sigma\nu}$
\\[0.2 cm]\hline
$ \ $&$\omega_{\mu\nu} :=
\gamma^{\sigma}_{\mu\alpha}\gamma^{\alpha}_{\nu\sigma}$
\\[0.2 cm]\hline
$ \ $&$\alpha_{\mu\nu} := C_{\mu}C_{\nu}$
\\[0.2 cm]\hline
\end{tabular}}
\end{center}}
\vspace{8pt}


\section{Semi-symmetric canonical connection}

A linear connection on $M$ is said to be semi-symmetric (\cite{r57}, \cite{AMR}, \cite{Ebtsam thesis}) if its torsion tensor $T^{\alpha}_{\mu\nu}$ is written in the form $T^{\alpha}_{\mu\nu} = \delta^{\alpha}_{\mu} \,\omega_{\nu} - \delta^{\alpha}_{\nu}\, \omega_{\mu},$ where $\omega_{\gamma}$ are the components of an arbitrary differential 1-form $\omega$.


Let $M$ be an AP-space with paralellization vector fields $\,\undersym{\lambda}{i}$ and metric $g$ defined by (\ref{metric AP}). Hence, $(M,\,g)$ can be considered as a Riemannian space. \textbf{We assume that the canonical connection $\Gamma^{\alpha}_{\mu\nu}$ of the AP-space,  given by (\ref{canonical}), is semi-symmetric}. An AP-space whose canonical connection is semi-symmetric will be referred to as an SAP-space.

From now on, we will be placed on an SAP-space $(M,\, \undersym{\lambda)}{i}$. Hence, the torsion tensor (\ref{tor}) is written in the form
\begin{equation}\label{localss}
 \Lambda^{\alpha}_{\mu\nu} = \delta^{\alpha}_{\mu} \,\omega_{\nu} - \delta^{\alpha}_{\nu}\, \omega_{\mu},
 \end{equation}
where $\omega_{\gamma}$ are the components of an arbitrary scalar form $\omega$. Consequently, the basic form (\ref{basic}) is given by
\begin{equation}\label{c&w}
                 C_{\mu}   =  (1 - n) \, \omega_{\mu}.\end{equation}
Moreover, as the canonical connection is metric, it can be written in the form \cite{r57}
\begin{equation}\label{SScanonical}
  \Gamma^{\alpha}_{\mu\nu} = \,\overcirc{\Gamma}{^{\alpha}_{\mu \nu}} + \delta^{\alpha}_{\mu}\, \omega_{\nu} - g_{\mu \nu}\, \omega^{\alpha},
   \end{equation}
where $\,\overcirc{\Gamma}{^{\alpha}_{\mu \nu}}$ is the Riemannian connection (\ref{Riemannian AP}) and $\omega^{\alpha} := g^{\alpha \beta}\, \omega_{\beta}.$
Consequently, the contortion tensor (\ref{cont}) is given by
 \begin{equation}\label{contortion S.S}
 \gamma^{\alpha}_{\mu \nu} = \delta^{\alpha}_{\mu}\, \omega_{\nu} - g_{\mu \nu}\, \omega^{\alpha}.\end{equation}
 \par
As the curvature tensor of the canonical connection vanishes, using (\ref{SScanonical}), the Riemannian curvature tensor\, $\overcirc{R}{^{\alpha}_{\mu\nu\sigma}}$ of\,\, $ \overcirc{\Gamma}^\alpha_{\mu\nu}$ is given by
$$\overcirc{R}{^{\alpha}_{\mu\nu\sigma}} = \delta^{\alpha}_{\mu}
(\omega_{\nu|\sigma} - \omega_{\sigma|\nu})
+ (g_{\mu\sigma}\, \omega^{\alpha} \ _{|\nu} - g_{\mu\nu} \,\omega^{\alpha} \
_{|\sigma})  + 2\omega^{\alpha}(g_{\mu\sigma}\,\omega_{\nu} -
g_{\mu\nu}\,\omega_{\sigma}).$$
Let $\,\overcirc{R}{_{\mu\nu}}:=\, \overcirc{R}{^{\alpha}_{\mu\nu\alpha}}$ be the Ricci tensor, then
\begin{equation}\label{the Ricci Riemannian curvature} \overcirc{R}{_{\mu\nu}} = \omega_{\nu|\mu} - g_{\mu\nu}\, \omega^{\sigma}\,_{|\sigma} +
2(\omega_{\nu}\,\omega_{\mu} - g_{\mu\nu}\,\omega^{\sigma}\,\omega_{\sigma}).\end{equation}
Consequently, the scalar curvature $\,\overcirc{R} \, := g^{\mu \nu}\, \overcirc{R}{_{\mu \nu}}.$ is given by
\begin{equation}\label{the scalar Riemannian curvature}  \overcirc{R} = (1 - n)({\omega^{\mu}\ _{|\mu}} + 2{\omega^{\mu}\omega_{\mu}}).\end{equation}
We have the following simple, interesting and unexpected result.
\begin{thm}\label{The second order covariant tensor}
The second order covariant tensor $\omega_{\nu|\mu}$ is symmetric:
\begin{equation}\label{der.commutes} \omega_{\mu|\nu} = \omega_{\nu|\mu}.\end{equation}
\end{thm}
\noindent The proof follows directly from (\ref{the Ricci Riemannian curvature}) and the fact that $\,\overcirc{R}{_{\mu\nu}}$ is symmetric.
\begin{rem}
\em{In the context of AP-geometry, most of the geometric objects are expressed in terms of the torsion tensor $\Lambda^{\alpha}_{\mu\nu}$. By assuming that the canonical connection is semi-symmetric, it is found that most of the geometric objects are expressed in terms of the basic vector $C_{\mu}$ or the $1$-form $\omega_{\mu}$} (via (\ref{c&w})).
\end{rem}


\section{Generalized Field Theory}

The construction of a purely geometric theory unifying gravity and
electromagnetism, the Generalized Field Theory (GFT), was successfully
established by Mikhail and Wanas in 1977 \cite{Mikhail}. The GFT is formulated in
the context of AP-geometry. The sixteen degrees of freedom
of AP-geometry (in dimension four) make this geometry suitable
for describing the gravitational field, which needs ten degrees of
freedom, in addition to the electromagnetic field, which needs six
degrees of freedom.

\vspace{5pt}
We give here a brief outline of the GFT. For more details, we refer to \cite{Mikhail}. Beginning with the Lagrangian density
\begin{equation*}\label{Lagrangian}\mathcal{L}: = \det(\lambda)\,g^{\mu\nu}L_{\mu\nu} := \det(\lambda)\,g^{\mu\nu}(\Lambda^{\alpha}_{\epsilon\mu}\Lambda^{\epsilon}_{\alpha\nu} - C_{\mu}C_{\nu}),
\end{equation*}
where $\Lambda^{\alpha}_{\mu\nu}$ and $C_{\mu}$ are given by $(\ref{tor})$ and $(\ref{basic})$ and $\det(\lambda)$ denotes the determinant of the matrix $(\,\undersym{\lambda}{i}_{\alpha})$,
Mikhail and Wanas, using a certain variational technique, obtained the differential identity
\begin{equation*}
E^{\mu}\!\,_{\nu\widetilde{|}\mu} = 0.
\end{equation*}

Considering this identity as representing  a certain conservation law, the field equations of the GFT are taken to be
\begin{equation}
\begin{split}\label{Vxx} E_{\mu\nu} :&= g_{\mu\nu}L - 2L_{\mu\nu} -
2(C_{\mu}C_{\nu} - C_{\nu|\mu}) + 2g_{\mu\nu}(C^{\epsilon}C_{\epsilon} - C^{\epsilon}\!\,_{|\epsilon})\\&
\ \ \ \, - \, 2(C^{\epsilon}\Lambda_{\mu\epsilon\nu} + g^{\epsilon\alpha}\Lambda_{\mu\nu\alpha|\epsilon})=0.
\end{split}
\end{equation}

The symmetric part of $(\ref{Vxx})$, can be written in the form
\begin{equation}\label{intx} R_{\mu\nu} - \frac{1}{2}\,g_{\mu\nu} R = T_{\mu\nu},\end{equation}
which may be considered as the Einstein field equations, where the energy-momentum tensor $T_{\mu\nu}$ is expressed in terms of the fundamental symmetric tensor fields of Table~2. Moreover, according to $(\ref{intx})$, $T_{\mu\nu}$ satisfies the conservation law
\begin{equation*}
T^{\mu\nu}\!\,_{;\,\mu} = 0.
\end{equation*}

On the other hand, the skew-symmetric part of (\ref{Vxx}) can be written in the form
\begin{equation}\label{curl}F_{\mu\nu} = C_{\mu, \nu} - C_{\nu, \mu}\end{equation}
where $F_{\mu\nu}$ is expressed in terms of the fundamental skew-symmetric tensor fields of Table 2. $F_{\mu\nu}$ may be interpreted as the electromagnetic field expressed as the curl of the basic form $C_{\mu}$.
In view of $(\ref{curl})$, $F_{\mu\nu}$ satisfies the (generalized) second Maxwell's equation
\begin{equation*}
{\mathfrak{S}_{\mu\nu\sigma}}\{ F_{\mu\nu;\,\sigma}\} ={\mathfrak{S}_{\mu\nu\sigma}}\{F_{\mu\nu, \sigma}\} = 0.
\end{equation*}
It should be noted that, in general, the gravitational and electromagnetic fields are not splitted completely unless we go to low-energy (week field) approximation.

\bigskip
To sum up, the field equations obtained are nonsymmetric. The symmetric
part of the field equations contains a second order tensor
representing the material distribution. This tensor is a pure
geometric, not a phenomenological, object. The skew-symmetric part of the
field equations gives rise to a generalized form of Maxwell's
equations in which the electromagnetic field is, again, purely
geometric. The skew-symmetric section of the theory is gauge invariant. The GFT coincides with both Maxwell's and Newton's
theories in the limits of weak static fields and slowly moving
test particles \cite{MW}. In the GFT, the metric tensor field $g_{\mu\nu}$ plays the role of the gravitational potential, while the basic form $C_{\mu}$ plays the role of the electromagnetic potential. Finally, all physical objects involved are expressed in terms of the fundamental tensor fields of the AP-space (Table 2).


\section{Field equations and Physical consequences}
\hspace{12pt}
We here construct field equations in the context of SAP-geometry. We take for the field equations a Lagrangian similar in form to that used in GFT (Section $3$). This is done for at least three reasons. First, the form of the chosen Lagrangian is relatively simple (depends on the vector fields $\lambda_{\beta}$ and their first derivatives $\lambda_{\beta,\gamma}$ {\it which are assumed to be independent}). Secondly, such form of the Lagrangian has led to powerful theoretical and experimental results in the context of AP-geometry (cf. \cite{MR}, \cite{**}, \cite{W1}, \cite{W2}). Finally, in order to facilitate comparison between our theory and the GFT.

\vspace{5pt}
Let M be an SAP-space with dimension $n \geq 3$\footnote{the reason for taking the dimension greater than two will be clarified later.}.
As easily checked, using $(\ref{canonical}),\,(\ref{tor})$ and $(\ref{localss})$, the following relation holds
\begin{equation}\label{yw}
\undersym{\lambda}{i}^{\alpha}\, \big(\,\undersym{\lambda}{i}_{\mu, \nu}-\undersym{\lambda}{i}_{\nu, \mu} \big) =\delta^{\alpha}_{\mu}\,\omega_{\nu}-\delta^{\alpha}_{\nu}\,\omega_{\mu}.
\end{equation}
Moreover, the curl of $\,\undersym{\lambda}{i}_{\mu}$ is thus given by
$$\,\undersym{\lambda}{i}_{\mu, \nu}-\undersym{\lambda}{i}_{\nu, \mu}=\,\undersym{\lambda}{i}_{\mu}\,\omega_{\nu}-\,\undersym{\lambda}{i}_{\nu}\,\omega_{\mu}.$$

\bigskip
Now, we start with the following scalar Lagrangian. Let
\begin{equation}\label{scalarl}
 \mathcal{H} := \det(\lambda) g^{\mu \nu} H_{\mu \nu},
 \end{equation}
where $\det(\lambda)$ denotes the determinant of the matrix $(\,\undersym{\lambda}{i}^{\alpha})$\, and
$$H_{\mu \nu} := \Lambda^{\alpha}_{\epsilon \mu} \,\, \Lambda^{\epsilon}_{\alpha \nu} - C_{\mu}  \,\, C_{\nu}$$

We assume that $\,\undersym{\lambda}{i}_{\beta}$ and $\,\undersym{\lambda}{i}_{\beta,\gamma}$ are independent. The Euler-Lagrange equations corresponding to $(\ref{scalarl})$ are given by
\begin{equation}\label{EL}
\frac{\delta \mathcal{H}}{\delta\,\undersym{\lambda}{i}_{\beta}} := \frac{\partial \mathcal{H}}{\partial\,\undersym{\lambda}{i}_{\beta}} - \frac{\partial}{\partial x^{\gamma}} \left( \frac{\partial \mathcal{H}}{\partial\,\undersym{\lambda}{i}_{\beta,\gamma}}\right) = 0.
\end{equation}
In view of $(\ref{localss})$ and $(\ref{c&w})$, we have
\begin{equation}\label{Homega}
    \mathcal{H} = (n-1)(2-n)\, det(\lambda)\, \omega^{2},
              \end{equation}where
    $$ \omega^{2} = g^{\mu \nu}\, \omega_{\mu} \,\omega_{\nu} = \omega^{\nu} \,\omega_{\nu}.   $$
It is clear from $(\ref{Homega})$ that $n=1$ or $n=2$ give the trivial result that $\mathcal{H}$ vanishes identically. It is for this reason that we take $n\geq3$.

\vspace{5pt}
Using $(\ref{yw})$,  $\omega_{\mu}$ can be expressed explicitly in terms of the $\lambda$'s in the form
\begin{equation}\label{omega is fn. on landa}
(1-n)\,\omega_{\mu}= \,\undersym{\lambda}{i}^{\alpha}\, \big(\,\undersym{\lambda}{i}_{\mu, \alpha}-\undersym{\lambda}{i}_{\alpha, \mu} \big).
 \end{equation}
Relation $(\ref{omega is fn. on landa})$ is interesting as it represents a strong link between the AP-structure and our imposed condition. It should be noted that we have started with $\omega$ arbitrary but the AP-context forced $\omega$ to be a function of the parallelization vector fields $\undersym{\lambda}{i}$.

\vspace{5pt}
We now evaluate the constituents of the Euler-Lagrange equations $(\ref{EL})$. To accomplish this we need the following lemma.

\begin{lem}\label{1}
Let M be an SAP-space with dimension $n \geq 3$. Then the following identities hold:
\begin{description}
\item[(a)] $\displaystyle \frac{\partial\det(\lambda)}{\partial\,\undersym{\lambda}{i}_{\beta}} = \,\undersym{\lambda}{i}^{\beta} \det(\lambda)$.
\item[(b)] $\displaystyle \dd \frac{\partial\det(\lambda)}{\partial\,\undersym{\lambda}{i}_{\beta , \gamma }} = 0$.
\item[(c)] $\displaystyle \dd \frac{\partial\det(\lambda)}{\partial\, x^{\gamma}} = \det(\lambda)\,\, \undersym{\lambda}{k}^{\mu} \,\, \undersym{\lambda}{k}_{\mu , \gamma}$.
\item[(d)] $\displaystyle \frac{\partial \omega^{2}}{\partial\,\undersym{\lambda}{i}_{\beta}}= \frac{2}{1-n}\,\Big[\,\undersym{\lambda}{i}^{\beta}\,\omega^{2}+(n-2)\, \,\undersym{\lambda}{i}^{\alpha}\, \omega_{\alpha}\, \omega^{\beta}  \Big]$.
\item[(e)] $\displaystyle \dd \frac{\partial\omega^{2}}{\partial\,\undersym{\lambda}{i}_{\beta , \gamma}} = \frac{2}{(1-n)} \, \Big[\,\undersym{\lambda}{i}^{\gamma}\, \omega^{\beta} - \omega^{\gamma}\, \undersym{\lambda}{i}^{\beta}\Big]$.
\end{description}
\end{lem}

\vspace{8pt}
Using (\ref{Homega}) and Lemma \ref{1}, we get:

\begin{lem}\label{2}
Let M be an SAP-space with dimension $n \geq 3$. Then the following identities hold:
\begin{description}
\item[(a)] $\displaystyle \frac{\partial \mathcal{H}}{\partial\,\undersym{\lambda}{i}_{\beta}} =  (n-2) \,\det(\lambda)\, \left[ 2(n-2)\, \undersym{\lambda}{i}^{\alpha}\, \omega_{\alpha} \, \omega^{\beta}-(n-3)\,  \undersym{\lambda}{i}^{\beta}\, \omega^{2} \right]$.
\item[(b)] $\displaystyle  \frac{\partial \mathcal{H}}{\partial\, \undersym{\lambda}{i}_{\beta , \gamma}} = 2(n-2) \, \det(\lambda) \, \big [\,\undersym{\lambda}{i}^{\gamma}\, \omega^{\beta} - \omega^{\gamma} \, \undersym{\lambda}{i}^{\beta} \big].$
\item[(c)] $\displaystyle \dd \frac{\partial}{\partial x^{\gamma}} \Big( \frac{\partial \mathcal{H}}{\partial\, \undersym{\lambda}{i}_{\beta , \gamma}} \Big) = 2\,(n-2)\, \det(\lambda)
\Big[\,\undersym{\lambda}{k}^{\mu}\, \undersym{\lambda}{k}_{\mu , \gamma}\,(\,\undersym{\lambda}{i}^{\gamma}\, \omega^{\beta} - \,\undersym{\lambda}{i}^{\beta}
\, \omega^{\gamma}) + \,\undersym{{{\lambda}^{\gamma}}_{,\gamma}}{i}\,\omega^{\beta} +\, \undersym{\lambda}{i}^{\gamma}\, {\omega^{\beta}}_{,\gamma}\nonumber\\ \phantom{mmmmmmn}- \undersym{{{\lambda}^{\beta}}_{,\gamma}}{i} \,\omega^{\gamma} - \,\undersym{\lambda}{i}^{\beta}\, {\omega^{\gamma}}_{,\gamma} \Big]$.
\end{description}
\end{lem}

\vspace{8pt}
Now, let us define the geometric object
\begin{equation}\label{E}
   E^{\beta}_{\sigma}:=  \frac{1}{det(\lambda)} \left(\frac{\delta \mathcal{H}}{\delta \,\undersym{\lambda}{i}_{\beta}} \right) \,\undersym{\lambda}{i}_{\sigma}.
 \end{equation}
Substituting the formulae of Lemma \ref{2} into (\ref{E}), using the definition (\ref{EL}) of $\dd \frac{\delta \mathcal{H}}{\delta\,\undersym{\lambda}{i}_{\beta}}$, we get
\begin{eqnarray*}\label{11075}                                                                                                                                                \dd E_{\sigma}^{\beta}
&=&(n-2) \, \Big[ (n+3)\, \delta^{\beta}_{\sigma}\, \omega^{2}+2\,\delta^{\beta}_{\sigma}\,{\omega^{\gamma}}_{|\gamma}-2{\omega^{\beta}}_{|\sigma} \Big].
\end{eqnarray*}
The tensor character of $E_{\sigma}^{\beta}$ is clear. Lowering the index $\beta$ of $E^{\beta}_{\sigma}$, we obtain the field equations $E_{\mu\nu}=0$, where
\begin{equation}\label{11077}                                                                                                                                                \dd E_{\mu\nu}= (n-2)\, \Big[(n+3)\, g_{\mu\nu}\,\omega^{2}
+2\,g_{\mu\nu}\,{\omega^{\gamma}}_{|\gamma}-2\,\omega_{\mu|\nu}\Big]. \end{equation}
By theorem \ref{The second order covariant tensor}, $\,\omega_{\mu|\nu}=\omega_{\nu|\mu}$,  $\,E_{\mu\nu}$ is symmetric.

Clearly, by $(\ref{the Ricci Riemannian curvature})$ and $(\ref{the scalar Riemannian curvature})$, the Einstein tensor $G_{\mu\nu}:=\,\overcirc{R}{_{\mu\nu}}- \frac{1}{2}\, g_{\mu\nu}\,\overcirc{R}$ can be expressed in terms of $\omega_{\mu}$ in the form
\begin{equation}\label{Gmunu}
G_{\mu\nu}=\omega_{\nu|\mu}+ \frac{n-3}{2}\,g_{\mu\nu}\,{\omega^{\gamma}}_{|\gamma}+ (n-3)\,g_{\mu\nu}\,\omega^{2}+ 2\,\omega_{\mu}\,\omega_{\nu}.\end{equation}
Taking $(\ref{Gmunu})$ into account, $(\ref{11077})$ takes the form
\begin{equation}\label{Ewithwn1}
\frac{1}{n-2}\,E_{\mu\nu}=G_{\mu\nu}-2\,\omega_{\nu}\,\omega_{\mu}+6\,g_{\mu\nu}\,\omega^{2}-3\,\omega_{\mu|\nu}-\frac{n-7}{2}\,g_{\mu\nu}\,
{\omega^{\gamma}}_{|\gamma}.
\end{equation}
In view of $(\ref{Ewithwn1})$, the field equations $E_{\mu\nu}=0$ give rise to
\begin{equation}\label{EwithCn2}
 \qquad \qquad \qquad\,\overcirc{R}{_{\mu\nu}}- \,\frac{1}{2}\, g_{\mu\nu}\,\overcirc{R}\,=2\,\omega_{\nu}\,\omega_{\mu}-6\,g_{\mu\nu}\,\omega^{2}+3\,\omega_{\mu|\nu}+\frac{n-7}{2}\,g_{\mu\nu}\,
{\omega^{\gamma}}_{|\gamma}. \end{equation}

If we set
\begin{equation}\label{Tmunu}
T_{\mu\nu} :=2\,\omega_{\nu}\,\omega_{\mu}-6\,g_{\mu\nu}\,\omega^{2}+3\,\omega_{\mu|\nu}+\frac{n-7}{2}\,g_{\mu\nu}\,
{\omega^{\gamma}}_{|\gamma}, \end{equation}
then, (\ref{EwithCn2}) takes the form
\begin{equation}\label{Einsteineq}
 \,\overcirc{R}{_{\mu\nu}}-\, \frac{1}{2}\, g_{\mu\nu}\,\overcirc{R}=T_{\mu\nu}. \end{equation}
By taking into account the divergence of $(\ref{Einsteineq})$, one can see, from the second Bianchi identity, that $T_{\mu\nu}$ satisfy the equations
\begin{equation}\label{dif.Tmunu}
 {T^{\mu\nu}}_{;\,\mu}=0.
\end{equation}
Hence, $T_{\mu\nu}$ can be interpreted as the energy-momentum tensor and (\ref{dif.Tmunu}) represents a conservation law. Unlike, the Einstein's field equations, the energy-momentum tensor defined by (\ref{Tmunu}) is purely geometric (Equations (\ref{Gmunu}) and (\ref{Tmunu}) show that both Einstein tensor $G_{\mu\nu}$ and the energy-momentum tensor $T_{\mu\nu}$ are expressed solely in terms of the geometric objects $g_{\mu\nu}$ and $\omega_{\mu}$). Furthermore, the gravitational potential can be attributed to the metric tensor $g_{\mu\nu}$.

\vspace{5pt}
The electromagnetic field strength is given by
\begin{equation*}
\label{emfs} F_{\mu\nu}= C_{\mu,\nu}-C_{\nu,\mu},
\end{equation*}
where $C_{\mu}$ is the electromagnetic potential. In view of (\ref{localss}), (\ref{c&w}) and (\ref{der.commutes}), we obtain
\begin{equation*}
F_{\mu\nu} \equiv 0.
\end{equation*}
This automatically implies that the electromagnetic field, which is represented in GFT by the skew symmetric part of the field equations $E_{\mu\nu}=0$, vanish identically.

\begin{rem}
\em{It is well known that if a linear connection (with covariant derivative~$||$) is symmetric, then $A_{\mu||\nu}-A_{\nu||\mu}=A_{\mu,\nu}-A_{\nu,\mu}$. It is interesting that such a relation holds here ($C_{\mu|\nu}-C_{\nu|\mu}=C_{\mu,\nu}-C_{\nu,\mu}$) though the canonical connection\, $\Gamma^{\alpha}_{\mu\nu}$\, is non-symmetric. This is due to the semi-symmetry condition.}
\end{rem}
We will refer to our theory, studying the GFT-Lagrangian in the SAP-context, as Special Generalized Field Theory (SGFT).


\section{A comparative study}

The fundamental second order tensors listed in Table 2 are necessary for physical applications. They can be used to determine what type of physical phenomena the geometry can describe (cf., for example, \cite{MW}).
\vspace{5pt}
\par
In our SGFT all fundamental skew-symmetric tensors vanish, namely,
$\xi_{\mu\nu},\, \gamma_{\mu\nu},$
$\,\eta_{\mu\nu},\,\chi_{\mu\nu},\,\epsilon_{\mu\nu}$ and $\kappa_{\mu\nu}$. For example, we have just shown that $\epsilon_{\mu\nu}:=C_{\mu|\nu}-C_{\nu|\mu}$ vanish. Also,
$\eta_{\mu\nu} := C_{\alpha}\,\Lambda^{\alpha}_{\mu\nu}= (1-n)\,\omega_{\alpha}\,(\delta^{\alpha}_{\mu}\,\omega_{\nu}-\delta^{\alpha}_{\nu}\,\omega_{\mu})=(1-n)(\omega_{\mu}\,\omega_{\nu}-\omega_{\nu}\omega_{\mu})
=0,$
and similarly for the other tensors.
\par
The vanishing of the fundamental skew symmetric tensors makes us sure that the resulting field equations of the SGFT, even before starting calculations, describe gravity only.

Table 3 compares between the fundamental symmetric tensors in AP-geometry and SAP-geometry. These tensors take a much simpler form thanks to $(\ref{c&w})$ and $(\ref{contortion S.S})$.

\begin{center} { Table 3: Fundamental symmetric second order tensors}\\[0.18cm]
{
\begin{tabular}{|c|c|}\hline
{AP-geometry}&{SAP-geometry}\\[0.2cm]\hline
$\sigma_{\mu\nu}=\gamma^{\alpha}_{\sigma\mu}\,\gamma^{\sigma}_{\alpha\nu} $& $\sigma_{\mu\nu}=(n-1)\, \omega_{\mu}\,\omega_{\nu}$\\[0.3cm]\hline
  $\omega_{\mu\nu}= \gamma^{\alpha}_{\mu\sigma}\,\gamma^{\sigma}_{\nu\alpha}$ & $\omega_{\mu\nu}=2(\omega_{\mu}\,\omega_{\nu}-g_{\mu\nu}\,\omega^{2})$\\[0.3cm]\hline
  $\alpha_{\mu\nu}= C_{\mu}\,C_{\nu}$ & $\alpha_{\mu\nu}= (n-1)\,\sigma_{\mu\nu}$\\[0.2cm]\hline
  $\theta_{\mu\nu}=C_{\mu|\nu}+C_{\nu|\mu}$ & $\theta_{\mu\nu}=2\,(n-1)\,\omega_{\mu|\nu}$\\[0.3cm]\hline
  $\psi_{\mu\nu}=\gamma^{\alpha}_{\mu\nu|\alpha}+\gamma^{\alpha}_{\nu\mu|\alpha} $ & $\psi_{\mu\nu}= 2\,(\omega_{\mu|\nu}-2\,g_{\mu\nu}\,{\omega^{\alpha}}_{|\alpha})$\\[0.3cm]\hline
 $\phi_{\mu\nu}=C_{\alpha}\,(\gamma^{\alpha}_{\mu\nu}+\gamma^{\alpha}_{\nu\mu})$  & $\phi_{\mu\nu}=(n-1)\,\omega_{\mu\nu}$ \\[0.3cm]\hline
  $\varpi_{\mu\nu}= \gamma^{\alpha}_{\mu\sigma}\,\gamma^{\sigma}_{\alpha\nu}+\gamma^{\alpha}_{\nu\sigma}\,\gamma^{\sigma}_{\alpha\mu}$ & $\varpi_{\mu\nu}=\omega_{\mu\nu}$ \\[0.3cm]\hline
\end{tabular}}
\end{center}

\smallskip
Using the above table, we can write $(\ref{Tmunu})$ in terms of the fundamental symmetric second order tensors as follows:
\begin{equation}\label{T2}
 T_{\mu\nu}=3\,\omega_{\mu\nu}- \frac{n-7}{8}\,\psi_{\mu\nu}- \frac{4}{n-1}\,\sigma_{\mu\nu}+ \frac{n+5}{8(n-1)}\,\theta_{\mu\nu}.
 \end{equation}
Moreover, the cosmological function defined by
\begin{equation*}
\Lambda:= \frac{1}{2}(\sigma-\omega),
\end{equation*}
where $\sigma:= g^{\mu\nu}\,\sigma_{\mu\nu}$ and $\omega:= g^{\mu\nu}\,\omega_{\mu\nu}$,
has the form
\begin{equation*}\label{cosm}
\Lambda= \frac{3}{2}(n-1)\,\omega^{2}.
\end{equation*}
It should be noted that our energy-momentum tensor $T_{\mu\nu}$ and cosmological function $\Lambda$ take a mush simpler form compared with those of the GFT. Moreover, the last equation implies that $\Lambda$ does not vanish.

\vspace{5pt}
Table 4 summarizes the most important results obtained so far.

\newpage
\begin{center} { Table 4: Comparison between SAP-geometry and AP-geometry}\\[0.25cm]
{\begin{tabular}{|c|c|}
  \hline
{AP-geometry}&{SAP-geometry}\\[0.25cm]\hline
  The most important tensor is $\Lambda^{\alpha}_{\mu\nu}$ &  The most important tensor is $C_{\mu}$ \\[0.18cm]\hline
   $C_{\mu}=0 \centernot\Longrightarrow \Lambda^{\alpha}_{\mu\nu}=0$& $C_{\mu}=0 \Longrightarrow \Lambda^{\alpha}_{\mu\nu}=0$ \\[0.18cm]\hline
   $C_{\epsilon}\,\Lambda^{\epsilon}_{\mu\nu} \neq 0$ &$C_{\epsilon}\,\Lambda^{\epsilon}_{\mu\nu} = 0 $  \\[0.18cm]\hline
   $C_{\mu|\nu}- C_{\nu|\mu}=C_{\mu,\nu}- C_{\nu,\mu}+C_{\epsilon}\,\Lambda^{\epsilon}_{\mu\nu}$&$C_{\mu|\nu}- C_{\nu|\mu}=C_{\mu,\nu}- C_{\nu,\mu}=0$ \\[0.18cm] \hline
     For $\alpha \neq \mu$ and $\alpha \neq \nu$ simultaneously,&For all $\alpha \neq \mu$ and $\alpha \neq \nu$ simultaneously, \\[0.18cm]
    $\Lambda^{\alpha}_{\mu\nu} \neq 0$ in general&$\Lambda^{\alpha}_{\mu\nu} = 0$  \\[0.18cm] \hline
   Fundamental skew-symmetric tensors & Fundamental skew-symmetric tensors   \\[0.1cm]
  do not vanish &  vanish \\[0.18cm] \hline
 $E_{\mu\nu}=0$ describe & $E_{\mu\nu}=0$  describe\\[0.1cm]
 gravity and electromagnetism &gravity only \\[0.18cm] \hline
\end{tabular}}
\end{center}

\vspace{6pt}
We end this section with Table 5 presenting a comparison between GR, GFT and SGFT.

\begin{center} {Table 5: Comparison between GR, GFT and SGFT}\\[0.25cm]
{\begin{tabular}
{|c|c|c|c|c|c|c|c|c|c|c|c|}\hline
 {Field} & { Field}    & {No. of}   & {Field}    & {Differential}\\
 {Theory}& { Variables}& {Field}    & {Equations}& {Identities}\\
         &             & {Variables}&            & \\ [0.2cm]\hline
{ GR}&$g_{\mu\nu}$&  $10$ &$G^{\mu}\,_{\nu} = 0$&$G^{\mu}\,_{;\,\mu} = 0$\\[0.2 cm]\hline
{ GFT}&$\,\undersym{\lambda}{i}^{\mu}$&$16$&$E^{\mu}\,_{\nu} = 0$&$E^{\mu}\,_{\nu\widetilde{|}\mu} = 0$\\[0.2cm]\hline
{SGFT}&$\,\undersym{\lambda}{i}^{\mu}$&$16$&$E^{\mu}\,_{\nu} = 0$&$ T^{\mu\nu}\,_{;\,\mu} = 0$\\[0.2cm]\hline
\end{tabular}}
\end{center}


\vspace{0.2cm}
\section{Concluding remarks}
\begin{itemize}
\item[$\bullet$]
    In this work, we consider an AP-space in which the canonical connection is semi-symmetric. The field equations are constructed by applying a variational technique to a suitable Lagrangian defined in terms of the torsion and the basic form of the space.

\item[$\bullet$]
    In view of (\ref{Homega}), the explicit dependence of the scalar Lagrangian on $n$ trivializes the cases $n = 1$ and $n = 2$.
    Consequently, we consider only the values of $n \geq 3$. It should be noted that the values $n =1$ and $n = 2$ are not forbidden, but
    are excluded on the ground that they imply the vanishing of $H$.
    Accordingly, without loss of generality, we may (and do) assume that $n\geq 3$.

\item[$\bullet$]
    Equation $(\ref{omega is fn. on landa})$ tells us that the $1$-form $\omega_{\mu}$ can be expressed in terms of the vector fields forming the parallelization. This provides a strong link between the semi-symmetry condition and the AP-structure.

\item[$\bullet$]
    The vanishing of the basic form is equivalent to the vanishing of the torsion tensor, that is,
    $$C_{\mu}=0 \Longleftrightarrow \Lambda^{\alpha}_{\mu\nu}=0.$$
    The implication $ \Lambda^{\alpha}_{\mu\nu}=0 \Longrightarrow C_{\mu}=0$ holds trivially. The reverse implication follows directly from the fact that $C_{\mu}= (1-n)\,\omega_{\mu}$ and $\Lambda^{\alpha}_{\mu\nu}= \delta^{\alpha}_{\mu}\, \omega_{\nu}-\delta^{\alpha}_{\nu}\, \omega_{\mu}$. This implication is not true in the general case of an AP-structure.

\item[$\bullet$]
    Unlike AP-geometry, in which most of the geometric objects are expressed in terms of the torsion tenser, in the SAP-context, the basic form plays the role of the torsion tensor: most of the  geometric objects can be expressed in terms of the basic form.
    This is one of the reasons that geometric objects and geometric relations in the SAP-context
    acquire a simpler form than their counterparts in the AP-context. This fact is due to the simplicity of the basic form (contracted torsion) compared with the torsion tensor.

\item[$\bullet$]
    Table 4 is quite revealing and sheds light upon various properties that differentiate between SAP-geometry and AP-geometry. It shows that some geometric objects, unlike their counterparts in AP-geometry,
    vanish identically. In particular, the fundamental skew symmetric tensers, some torsion components and, last but not least,  the contraction of the basic form with the torsion tenser. Moreover, it gives an inverse implication that is not true in general.

\item[$\bullet$]
     A crucial difference between our field equations and those of the  GFT is that the Lagrangian in the GFT depends \emph{implicitly} on the dimension $n$ of the underlying manifold (being defined in terms of the $n$ vector fields forming the parallelization), whereas its counterpart in our field equations depends \emph{explicitly} on $n$.

\item[$\bullet$] Under the additional condition that the canonical connection is semi-symmetric, our chosen Lagrangian acquires a much simpler form compared to Mikhail-Wanas Lagrangian of the GFT. Moreover, this assumption gives rise to the interesting property that the field equations describing the electromagnetism disappear.
    This is because, as stated above, all fundamental skew-symmetric tensors vanish, a property which obviously does not hold in the AP-context.

\item[$\bullet$]
    Though the field equations under the semi-symmetry condition describe only gravity, it has a big advantage compared to Einstein's field equations. Unlike the classical general theory of relativity, the energy-momentum tensor $(\ref{Tmunu})$ (or (\ref{T2})) has a geometric origin. All geometric entities in our field equations are expressed in terms of both the metric $g_{\mu\nu}$ and $1$-form $\omega_{\mu}$. The field equations obtained may be used in physical applications related to inflation and the status of early universe.

\item[$\bullet$]
    One of the possible reasons why our field equations describe only gravity is that our imposed condition of semi-symmetry seems to be too strong, hence too restrictive. We conjecture that, by relaxing the semi-symmetry condition, namely, replacing $$\Lambda^{\alpha}_{\mu\nu}= \delta^{\alpha}_{\mu}\,\omega_{\nu}-\delta^{\alpha}_{\nu}\,\omega_{\mu}$$ by $$\Lambda^{\alpha}_{\mu\nu}= L^{\alpha}_{\mu}\,\omega_{\nu}-L^{\alpha}_{\nu}\,\omega_{\mu},$$
    where $L^{\alpha}_{\mu}$ is an arbitrary tensor field of type $(1,1)$, the field equations describing electromagnetism will not disappear. This point may be the subject of future research.

\item[$\bullet$] This work can be continued and extended. The equations of motion in the context of SAP-geometry can be studied. The field equations and the equations of motion under the assumption that the canonical connection is semi-symmetric can be studied in the more general context of GAP-geometry $\cite{GAP}$, EAP-geometry (\cite{amr}, \cite{quant}) or FP-geometry $\cite{Ebtsam}$.

\item[$\bullet$] Finally, in view of Theorem 1 of Yano \cite{r57}, the following result follows \cite{AMR}.
    If $(M,\,\undersym{\lambda}{i})$ is an SAP-space, then the associated Riemannian metric $g_{\mu\nu}= \undersym{\lambda}{i}_{\mu} \,\, \undersym{\lambda}{i}_{ \nu}$ is conformally flat. Some physical consequences may arise from such result. This needs more investigation.
\end{itemize}

\vspace{-1.5cm}
\providecommand{\bysame}{\leavevmode\hbox
to3em{\hrulefill}\thinspace}
\providecommand{\MR}{\relax\ifhmode\unskip\space\fi MR }
\providecommand{\MRhref}[2]{%
  \href{http://www.ams.org/mathscinet-getitem?mr=#1}{#2}
} \providecommand{\href}[2]{#2}


\begin{thebibliography}{10}

\bibitem{AE}
A. Einstein, \emph{ The meaning of relativity}, 5th. Ed., Princton Univ. Press, 1955.

\bibitem {FI}
F. I. Mikhail. {\it Tedrad vector fields and generalizing the theory of relativity},
Ain Shams Sc. Bull., {\bf 6} (1962), 87-111.

\bibitem{Mikhail}
F. I. Mikhail and M. I. Wanas, \emph{ A generalized field theory, I. Field equations},
Proc. R. Soc. London, A, {\bf 356} (1977), 471-481.

\bibitem{MW}
F. I. Mikhail and M. I. Wanas, \emph{A generalized field theory. II. Linearized field equations}, Int. J. theoret. phys., {\bf 20} (1981), 671.

\bibitem{MR}
A. Mozumder and D. Ray, \emph{Charged spherically symmetric solution in Mikhail-Wanas field theory}, Int. J. Theret. Phys., \textbf{29} (1990), 431-434.

\bibitem{Nashed1}
G. G. L. Nashed, \emph{Spherical symmetric charged-DS solution in $f(T)$-gravity theories}, Phys. Rev. D, \textbf{88} (2013), 104034.

\bibitem{Nashed2}
G. G. L. Nashed, \emph{A special exact spherical symmetric solution in $f(T)$-gravity theories}, Gen. Rel. Gravit., \textbf{45} (2013), 1887-1899.

\bibitem{Nashed3}
G. G. L. Nashed, \emph{Schwarzschild-NUT solutions in modified teleparallel gravity theory}, Astrophys. Space Sci., \textbf{350} (2014), 791-767.

\bibitem{Nashed4}
G. G. L. Nashed, \emph{Local Lorentz transformations and exact spherically symmetric vacuum solutions in $f(T)$-gravity theories}, Eur. Phys. J., C \textbf{73} (2014), 2394.

\bibitem{**}
R.~D.~S.~de Souza and R.~Opher,
\emph{Origin of $10^{15}-10^{16}$G magnetic fields in the central engine of gamma ray bursts,}
JCAP {\bf 1002}, 022 (2010). arXiv: 0910.5258 [astro-ph.HE].

\bibitem{Ebtsam thesis}
Ebtsam H. Taha, \emph{On the geometry of Finslerized parallelizable maifolds}, M. Sc. Thesis, Cairo Univeristy, 2013.

\bibitem{W}
M. I. Wanas, \emph{Absolute parallelism geometry: Developments, applications and problems}, Stud. Cercet. Stiin., Ser. Mat., \textbf{10} (2001), 297-309. arXiv: gr-qc/0209050.

\bibitem{W1}
M. I. Wanas and S. A. Ammar \emph{Spacetime structure and electromagnetism}, Mod. Phys. Lett. A, \textbf{25} (2010),1705-1721.
arXiv: gr-qc/0505092.

\bibitem{W2}
M. I. Wanas and S. A. Ammar \emph{A pure geometric approach to stellar structure}, Cent. Euro. J. Phys., \textbf{11} (2013), 936-948.
arXiv: 0906.4808 [gr-qc].

\bibitem{GAP}
M. I. Wanas, Nabil L. Youssef and A. M. Sid-Ahmed, {\it Geometry of parallelizable manifolds in the context of generalized Lagrange spaces}, Balkan J. Geom.  Appl., \textbf{13}, \textbf{ 2} (2008), 120-139. arXiv: 0704.2001 [gr-qc].

\bibitem{quant}
M. I. Wanas, Nabil L. Youssef and A. M. Sid-Ahmed, \emph{Teleparallel Lagrange geometry and a unified field theory}, Class. Quant. Grav., \textbf{27} (2010), 045005, 29 pages. arXiv: 0905.0209 [gr-qc].

\bibitem{r57}
K. Yano, \textit{On semi-symmetric metric connection}, Rev. Roumaine Math. Pures. Appl., \textbf{15} (1970), 1579-1586.

\bibitem{YE}
Nabil L. Youssef and Waleed A. Elsayed, \emph{A global approach to absolute parallelism geometry}, Rep. Math. Phys., \textbf{72} (2013), 1-23. arXiv: 1209.1379 [gr-qc].

\bibitem{AMR}
Nabil L. Youssef and A. M. Sid-Ahmed, {\it Linear connections and curvature tensors in the geometry of parallelizabl manifolds}, Rep. Math. Phys., \textbf{60}  (2007), 39-53. arXiv: gr-qc/0604111.

\bibitem{amr}
Nabil L. Youssef and A. M. Sid-Ahmed, {\it Extended absolute parallelism geometry}, Int. J. Geom. Meth. Mod. Phys., \textbf{5} (2008), 1109-1135. arXiv: 0805.1336 [math.DG].

\bibitem{Ebtsam}
Nabil L. Yousssef, A. M.~Sid-Ahmed and E. H. Taha, \emph{On Finslerized absolute parallelism spaces},  Int. J. Geom. Meth. Mod. Phys., \textbf{10}  (2013), 1350029 (21 pages). arXiv: 1206.4505 [math.DG].

\end{thebibliography}
 \end{document}